\documentclass[epj]{svjour}
\usepackage{graphics}
\usepackage{epsfig}
\begin{document}
\title{The theory and phenomenology of jets in nuclear collisions}
\author{Ivan Vitev\inst{1} \and Ben-Wei Zhang\inst{1,2} \and{Simon Wicks}\inst{3}  
}                     
\offprints{Ivan Vitev}          
\institute{Los Alamos National Laboratory, Theoretical Division,
MS B238, Los Alamos, NM 87545, USA  \and 
 Institute of Particle Physics, Hua-Zhong Normal University,
Wuhan, 430079, China   \and 
Department of Physics, Columbia University, 538 West 120-th Street,
New York, NY 10027, USA} 
\date{Received: date / Revised version: date}
%
\abstract{  
We review selected results from a recent in-depth study 
of jet shapes and jet cross sections in ultra-relativistic reactions 
with heavy nuclei at the LHC~\cite{VWZ}. We demonstrate that at the 
highest collider energies these observables become feasible as a new, 
differential and accurate test of the underlying  QCD theory. Our 
approach allows for detailed simulations of the experimental 
acceptance/cuts that help isolate jets emerging from a dense QGP.
We show for the first time that the pattern of stimulated
gluon emission can be correlated with a variable quenching 
of the jet rates and provide an approximately model-independent 
approach to determining  the characteristics of the medium-induced 
bremsstrahlung spectrum. The connection between such cross section 
attenuation and the in-medium jet shapes is elucidated.  
\PACS{
      {13.87.-a}{Jets in large-Q$^2$ scattering}   \and
      {24.85.+p}{Quarks, gluons, and QCD in nuclear reactions} \and
      {12.38.Mh}{Quark gluon plasma}
     } 
} 
\maketitle

\section{Introduction}
\label{intro}

At present, most measurements of hard processes in heavy ion collisions  
are limited to single particles and particle correlations, which are only 
the leading fragments of a jet. There is general agreement that final-state
parton energy loss in the QGP controls inclusive particle suppression,
the experimental methodology of determining $R_{AA}(p_T)$ or $I_{AA}(p_{T_1},p_{T_2})$
is well-established,  and the theory of jet quenching has been very 
successful~\cite{Gyulassy:2003mc}. The data, however, cannot resolve 
the staggering order of magnitude "systematic" discrepancy in the  medium 
properties extracted via competing phenomenological models~\cite{Adare:2008cg}. 
It is,  therefore, critical  to  find alternatives that  accurately reflect the 
energy  flow in strongly-interacting  systems and have a more direct connection 
to the underlying QCD theory.

The intra-jet energy distribution and the related cross
section for jets in the case of heavy ion reaction closely match the
criteria outlined above. In this paper we study the magnitude of 
in-medium  modification for these observables  in Pb+Pb collisions 
at $\sqrt{s}=5.5$~TeV at LHC. We demonstrate 
that a natural generalization of leading particle suppression to jets,
\begin{equation}
R_{AA}^{{\rm jet}}(E_T; R^{\max},\omega^{\min})  =
\frac{ \frac{d\sigma^{AA}(E_T;R^{\max},\omega^{\min})}{dy d^2 E_T} }
{ \langle  N_{\rm bin}  \rangle
\frac{d\sigma^{pp}(E_T;R^{\max},\omega^{\min})}{dy d^2 E_T}  } \; ,
\label{RAAjet}
\end{equation}
is sensitive to the nature of the medium-induced energy loss. The
steepness of the final-state differential spectra amplifies the observable effect
and the jet radius $R^{\max}$ and the minimum particle/tower energy
$p_{T\, \min} \approx \omega^{\min}$ provide, through the
evolution of $R_{AA}^{\rm {jet}}(E_T; R^{\max},\omega^{\min})$ at any fixed
centrality, experimental access to the QGP response to  quark and gluon
propagation.

We refer the reader to~\cite{Ellis:2007ib}
for a summary of the complications in defining a jet and the related topic
of jet-finding algorithms. To make our discussion simpler, we will assume
that the complications of the different definitions can be subsumed into
an $R_{sep}$ parameter~\cite{VWZ}. Once a jet axis and all of the jet
particles / calorimeter towers ``i'' have been identified, the ``integral jet     
shape''  is defined as:
\begin{equation}
 \Psi_{\rm {int}}(r;R) = \frac{\sum_i (E_T)_i \Theta (r-(R_{\rm {jet}})_i)}
{\sum_i  (E_T)_i  \Theta(R-(R_{\rm {jet}})_i)}
\end{equation}
where $r,R$ are Lorentz-invariant opening angles, $R_{ij} =                       
\sqrt{(\eta_i-\eta_j)^2 + (\phi_i-\phi_j)^2}$, and $i$ represents
a sum over all the particles in this jet. $\Psi_{\rm {int}}(r;R)$ is the
fraction of the total energy of a jet of radius $R$ within a sub-cone
of radius $r$.  It is automatically normalized so that
$\Psi_{\rm {int}}(R;R) = 1$ and the differential jet shape is defined as follows:
\begin{equation}
 \psi(r;R) = \frac{d\Psi_{\rm {int}}(r;R)}{dr}  \;.
\end{equation}
Understanding the many-body QCD theory behind jet shape calculations 
will naturally lead to understanding the attenuation of jets in 
reactions with heavy nuclei.

\section{Theoretical results for vacuum jet shapes}
Jet shapes in vacuum will provide the baseline for jet shape
studies in nuclear collisions. Here, we follow the methodology 
outlined in Ref.~~\cite{Seymour:1997kj} and generalize this approach
to include finite detector acceptance.

\subsection{Leading order results}

A fixed order QCD calculation of jet shapes is based on  the splitting 
functions $P_{a \rightarrow bc}(z)$, the
distributions of the large fractional lightcone momenta
(or approximately the energy fractions) of the fragments relative
to the parent parton, $z$ and $1-z$ respectively. To lowest order,
recalling that $\psi_a(r;R)$ describes the energy flow $\propto z$,
we can write:
\begin{equation}
   \psi_a(r;R)  =
\sum_b \frac{\alpha_s}{ 2 \pi }
\frac{2}{r} \int_{z_{min}}^{1-Z} dz  \,  z P_{a \rightarrow bc}(z) ,
\label{kernel}
\end{equation}
with $Z$ is defined as follows:
\begin{eqnarray}
  Z &=& \max \left\{ z_{min},   \frac{r}{r+R} \right\} \;  \mbox{ if }  
\;
r < (R_{sep} - 1)R  \;,   \\
   Z  &=& \max \left\{ z_{min},  \frac{r}{R_{sep}R} \right\} \;  
\mbox{ if } \;
r > (R_{sep} - 1)R  \; .
\label{Zpar}
\end{eqnarray}
In Eq.~(\ref{kernel})  $ r = (1-z) \rho $ is related to the opening
angle $\rho$ between the final-state partons.
In ``elementary'' $p+p$ collisions the inclusion of soft particles
($ z_{min} \approx 0 $) in theoretical calculations is not a bad
approximation. 
In heavy ion reactions, especially  for the most interesting case of  
central collisions, there is an enormous background of soft particles 
related to the bulk
QGP properties. Jet studies will likely require  minimum particle energy
$ > 1-2$~GeV at RHIC and even more stringent cuts at the LHC.
Furthermore, control over $z_{min}$ can provide detailed information
about the properties of QGP-induced bremsstrahlung. The kinematic
constraints on the values of $z$ are discussed in detail in~\cite{VWZ}.
We find for the jet shape functions for quarks and gluons:
\begin{eqnarray}
 && \psi_q(r) = \frac{C_F \alpha_s}{2\pi} \frac{2}{r}  \left(
  2 \log \frac{1-z_{min}}{Z}  \right. 
\left. - \frac{3}{2} \left[ (1-Z)^2 \right. \right. \nonumber 
\\ &&   \left. -z_{min}^2 \right]  \Big)  \; ,
\label{psiLO1} \\
 && \psi_g(r) = \frac{C_A \alpha_s}{2\pi} \frac{2}{r}
\left( 2 \log \frac{1-z_{min}}{Z}     \right. 
  \left. -  \left(  \frac{11}{6}
- \frac{Z}{3} + \frac{Z^2}{2} \right)   \right. \nonumber \\
&& \left.\times(1-Z)^2   + \left( 2 z_{min}^2 -\frac{2}{3} z_{min}^3
+ \frac{1}{2} z_{min}^4  \right) \right) \nonumber \\
     &&  + \frac{T_R N_f \alpha_s}{2\pi} \frac{2}{r}   \left(
  \left( \frac{2}{3} - \frac{2Z}{3} + Z^2 \right) (1-Z)^2 
 -\Big(  z_{min}^2 
 \right.
  \nonumber \\
&& \left. \left.  - \frac{4}{3} z_{min}^3 + z_{min}^4
\right)  \right) \;.
\label{psiLO2}
\end{eqnarray}
In the $z_{min} \rightarrow 0$ limit Eqs.~(\ref{psiLO1})   
and~(\ref{psiLO2})
reduce to previously known results~\cite{Seymour:1997kj}.

\begin{figure}[!t] 
\begin{center}
\epsfig{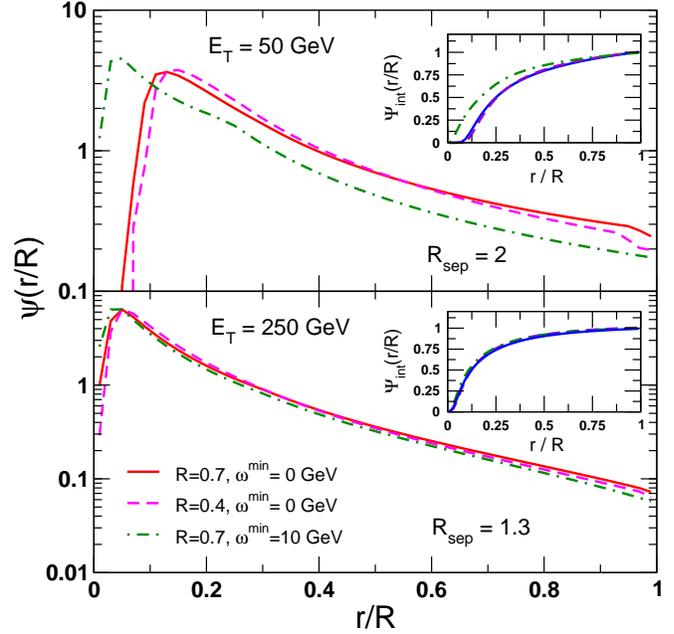}
\end{center}
\caption{ Numerical results for the differential jet shape 
in $p+p$ collisions at $\sqrt{s}=5.5$~TeV at the LHC. Solid lines 
represent jet shapes  with 
$R=0.7,\, \omega^{\min}=0$~GeV, dashed lines stand for jet shapes with 
$R=0.4,\; \omega^{\min}=0$~GeV, and dashed-dotted
lines are for jet shapes with $R=0.7,\; \omega^{\min}=10$~GeV. The inserts show
integrated jet shapes $\Psi_{\rm int}(r;R)$. 
}
\label{fig:LHCshapes}
\end{figure}

\subsection{The full vacuum  jet shape and numerical simulations}

A number of important effects beyond the fixed order, Eqs.~(\ref{psiLO1})   
and~(\ref{psiLO2}), contribute to the observed jet shape~\cite{VWZ,Seymour:1997kj}.
Taking all contributions  and ensuring that there is no double
counting at small $r/R$ to $ {\cal{O}}(\alpha_s)$ we find:
\begin{eqnarray}
\psi(r)&=&\psi_{\texttt{coll}}(r)\left( P(r)-1\right) +
\psi_{\texttt{LO}}(r) + \psi_{i,\texttt{LO}}(r)
\nonumber \\ && + \psi_{\texttt{PC}}(r)
+ \psi_{i,\texttt{PC}}(r) \; ,
\label{totpsi}
\end{eqnarray}
On the right-hand-side of Eq.~(\ref{totpsi}) the first term comes 
from  Sudakov resummation  with  subtraction of the leading  $1/r$,
$ (1/r) \log (1/r) $ contribution at small $r/R$ to avoid double
counting with the fixed order component of the differential jet shape. 
The second and  third terms represent the leading-order contributions in the 
final-state  and the initial-state.  The last two terms represent  the effect 
of  power corrections. In a full calculation the relative quark and gluon 
fractions $f_q+f_g = 1$  are also needed: $ \psi (r;E_T) = f_q(E_T,\sqrt{s})  
\psi_q(r;E_T) + f_g(E_T,\sqrt{s})   \psi_g(r;E_T)$. This theoretical model
is validated against CDF II data~\cite{VWZ}.

Figure~\ref{fig:LHCshapes}  shows our numerical results for the jet shape for
two different energies $E_T = 50$~GeV and $250$~GeV and two cone 
radii $R=0.7,\; 0.4$  in $p+p$ collisions at  $\sqrt{s}=5.5$~TeV at LHC. 
When plotted against the relative opening angle $r/R$,  these shapes are 
self-similar, i.e. approximately independent 
of the absolute cone radius $R$. One of the main theoretical developments
in Ref~\cite{VWZ} is the analytic approach to studying finite detector 
acceptance effects or experimentally imposed low momentum cuts. In
Fig.~\ref{fig:LHCshapes} this is illustrated via the selection of
$z_{min} = p_{T\, \min} / E_T = 0.2, \;  0.04$ ($p_{T\, \min} 
= \omega^{\min} = 10$~GeV). Eliminating the soft partons naturally 
leads to a  narrower  branching  pattern. However, for this effect to be 
readily observable $10-20\%$ of the jet energy, going into soft particles,
must be missed. Thus, even with $p_{T\, \min} \sim$few GeV cuts in Pb+Pb 
collisions at the LHC aimed at reducing or eliminating the background 
of bulk QGP particles that accidentally fall within the jet cone, the 
alteration of $\psi(r/R)$ is expected to be small. 
The integral jet shape $\Psi_{\rm int}(r/R)$, shown in the 
inserts of Fig.~\ref{fig:LHCshapes},  is not optimal as a tool for 
identifying kinematic and dynamic effects on jets due to its much smaller 
sensitivity.

\section{Medium-induced contribution to the  jet shape}
\label{sec:0}

The principal medium-induced contribution to a jet shape comes from
the radiation pattern of fast quarks or gluons, stimulated by their
propagation and interaction in the QGP. There is a simple heuristic
argument which allows one to understand how interference and coherence
effects in QCD amplify the difference between the energy distribution
in a vacuum jet and the in-medium jet shape~\cite{Vitev:2008jh}.
Any destructive effect on the integral average parton energy
loss  $\Delta E^{rad}$, such as the Landau-Pomeranchuk-Migdal effect,
can be traced at a differential level to the attenuation or full
cancellation  of the collinear,  $k_T \ll \omega$,  gluon  
bremsstrahlung:
\begin{eqnarray}
&& \Delta E^{rad}_{\rm LPM \; suppressed} \Rightarrow
\frac{dI^g}{d\omega}(\omega \sim E)_{\rm LPM \; suppressed} \nonumber \\
&& \Rightarrow  \frac{dI^g}{d\omega d^2 k_T}
(k_T \ll \omega )_{\rm LPM \; suppressed} \;, \;
\label{largeang}
\end{eqnarray}
and we indicate the parts of phase space where the modification of
the incoherent $ {dI^g} / {d\omega d^2 k_T} $ is most effective.

In our calculation we  use the GLV formalism of expanding the
medium-induced radiation in the correlations between multiple scattering
centers~\cite{GLV,Vitev:2008vk}. Theoretical interest in the angular bremsstrahlung 
distribution was stimulated by experimental measurements of enhanced
triggered opposite-side particle correlations away from the naive 
$\Delta \phi = \pi$~\cite{:2008cq,Ulery:2006ha,Molnar:2008jg}.
It has been know that the leading $ n = 1 $  contribution to
final-state medium-induced radiation does not have a collinear 
with the jet axis component~\cite{Vitev:2005yg}.
We now show~\cite{VWZ} that this result is general and holds to any order 
in the  opacity expansion: 
\begin{eqnarray}
\lim_{ r \rightarrow 0 } \frac{ \omega dN^g_{\rm med}}
{d\omega d \phi  d r }  = 0 \;.
\label{zero}
\end{eqnarray}
Numerical simulations, using  Monte-Carlo techniques,
confirm independently  that  ${dI^g}/{d\omega d^2 {\bf k}}$
vanishes as  ${\bf k} \rightarrow 0$  ~\cite{Wicks:2008ta}.

\begin{figure}[!t]
\begin{center}
\psfig{file=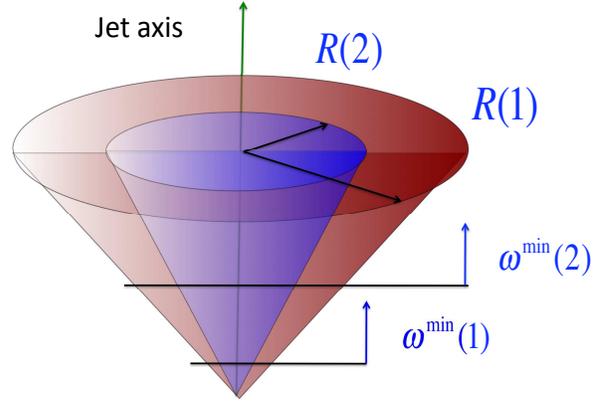,height=3.5in,width=2.8in,clip=,angle=-90}
\end{center}
\caption{  Schematic illustration of the cone radius  $R$
and the  particle/tower $p_T$ / $E_T$ selection. The measured energy is
the one that comes from particles with  $p_T > \omega_{min}$ and  
within $R$.
}
\label{fig:selectCone}
\end{figure}

The implication of our finding is that there is very little
overlap  between  the  techniques used to compute the ``vacuum'' and
medium-induced contributions to the jet shape. Also, selecting  
different jet radii $R$ and $p_{T\; \min}$ of  the particles will
significantly alter both the jet shape and the amount of energy lost  
by the hard
parton which can be recovered in the experimental measurement.
A clear strategy will  be to use the leverage arms
provided  by $R$ ($=R^{max}$  in the evaluation of the $\Delta E_{\rm  
rad}$) and $p_{T\, \min}$
($=\omega^{min}$  in the evaluation of the $\Delta E_{\rm rad}$) to  
determine the
distribution of the lost energy. This is illustrated schematically in
Fig.~\ref{fig:selectCone}. Theoretically, the first quantity to be  
calculated
is:
\begin{eqnarray}
  \frac{\Delta E^{in}}{E}(R^{\max},\omega^{\min})
= \frac{1}{E} \int_{\omega^{\min}}^E d\omega
  \int_0^{R^{\max} } dr  \frac{dI^g}{d\omega dr} (\omega,r) \; .
\nonumber \\
\label{def:out}
\end{eqnarray}

\begin{figure}
\center{\resizebox{0.42\textwidth}{!}{%
 \includegraphics{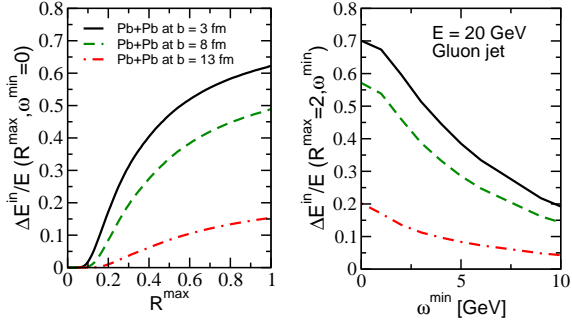}
} }\vspace*{.2cm} \caption{2D projections of Eq.~(\ref{def:out}).
The left panel shows the fractional energy loss dependence on the jet
radius $R^{\max}$ $(\omega^{\min} = 0)$
and the right panel shows this dependence versus the  acceptance cut 
$ \omega^{\min}$
($R^{\max}= R^{\infty}$). A gluon jet of $E_{jet} = 20$~GeV
in $b = 3, \; 8, \; 13$~fm   $Pb+Pb$ collisions at LHC was used as an  
example.  }
\label{fig:ERatio-2d}
\end{figure}

The separate dependence of ${\Delta E^{in}(R^{\max},\omega^{\min})}/{E} $ on
the cone radius and the momentum acceptance cut is  more clearly  
illustrated  in
Fig.~\ref{fig:ERatio-2d}. We show central, mid-central and peripheral  
collisions,
impact parameters $b=3, \; 8, \; 13$~fm, respectively, in $Pb+Pb$  
reactions
at LHC at nominal $\sqrt{s}$. We notice that, not surprisingly, the  
ratio
$\Delta E^{in}(R^{\max},\omega^{\min})/E$ goes down at larger impact  
parameters
because the energy loss of the jet decreases in peripheral collisions.
More importantly, at each impact parameter there is a variation of the  
amount
of the bremsstrahlung energy, recovered in the cone. This is precisely
the variation that will map on the $R_{AA}^{\texttt{jet}}(E_T, R^{\max}, 
\omega^{\min})$ observable.

\section{Tomography of jets in heavy ion collisions}
\label{sec:1}

The purpose of this Section is to relate the theory of jet propagation
in the QGP to experimentally measurable quantities.

\subsection{Experimental observables}
\label{sec:2}

An essential ingredient that controls the relative contribution of
$\psi_{\rm vac.}(r/R)$ and $\psi_{\rm med.}(r/R) = (1/\Delta E_{\rm rad}) dI^g/dr$ 
to the observed differential jet shape in heavy ion reactions and  also 
determines the attenuation of the jet cross sections is:
\begin{eqnarray}
f &\equiv& 
 \frac{\Delta E_{\rm rad}\left\{ (0,R);(\omega^{\min},E) \right\} }
{\Delta E_{\rm rad} \left\{ (0,R^\infty);(0,E) \right\} } \; ,
\end{eqnarray}
the {\em fraction} of the lost energy that falls within the jet cone, $r < R$, 
and carried by gluons of $\omega > \omega^{\min}$ relative to the 
total parton energy loss without the above kinematic constraints. If this 
fraction is known  together with the probability distribution $P(\epsilon)$
for the parton energy loss  the medium-modified jet cross section per 
binary $N+N$ scattering can be calculated as follows:
\begin{eqnarray}
\frac{\sigma^{AA}(R,\omega^{\min})}{d^2E_Tdy} &=& \int_{\epsilon=0}^1 
d\epsilon \; \sum_{q,g} P_{q,g}(\epsilon) 
\frac{1}{ (1 - (1-f_{q,g}) \cdot \epsilon)^2} \nonumber \\
&&\times
\frac{\sigma^{NN}_{q,g}(R,\omega^{\min})} {d^2E^\prime_Tdy} \; , 
\label{eq:sigma-AA}
\end{eqnarray}
where $E^\prime_T = E_T/(1 - (1-f_{q,g})\cdot \epsilon)$. The   
$(1-f_{q,g})\cdot \epsilon$  factor accounts for the total 
''missed'' energy in a jet cone measurement, which
necessitates $E^\prime_T > E_T$.
Next, we obtain the full jet shape, including the contributions 
from the vacuum and the medium-induced bremsstrahlung:
\begin{eqnarray}
\psi_{\rm tot.}\left({r}/{R}\right) &=&  
\frac{1}{\rm Norm}  \int_{\epsilon=0}^1 
d\epsilon \; \sum_{q,g} P_{q,g}(\epsilon) 
\frac{1}{ (1 - (1-f_{q,g}) \cdot \epsilon)^3} \nonumber \\
&&\times
\frac{\sigma^{NN}_{q,g}(R,\omega^{\min})} {d^2E^\prime_Tdy} 
\Big[ (1- \epsilon) \; 
\psi_{\rm vac.}^{q,g}\left({r}/{R}\right) 
\nonumber \\
&& \hspace*{2.7cm}  +  \, f_{q,g}\cdot \epsilon \; 
\psi_{\rm med.}^{q,g}\left(r/R\right) \Big] \; . \qquad
\label{psitotmed}
\end{eqnarray}  
We recall that, by definition, the area under any  differential
jet shape, $\psi_{\rm tot.}\left({r}/{R}\right)$, 
$\psi_{\rm vac.}\left({r}/{R}\right)$ and $\psi_{\rm med.}\left({r}/{R}\right)$,
is normalized to unity. Integrating over $r$ in Eq.~(\ref{psitotmed}),
we can easily see that the correct ``Norm'' is the quenched
cross section,  Eq.~(\ref{eq:sigma-AA}).  Discussion of
simple limiting cases for  Eqs.~(\ref{eq:sigma-AA})  and~(\ref{psitotmed})
which provide insight into the  contributions to the full 
jet shape and the jet cross section is given in~\cite{VWZ}.

\subsection{Numerical results}

\begin{figure}[!t]
\begin{center}
\epsfig{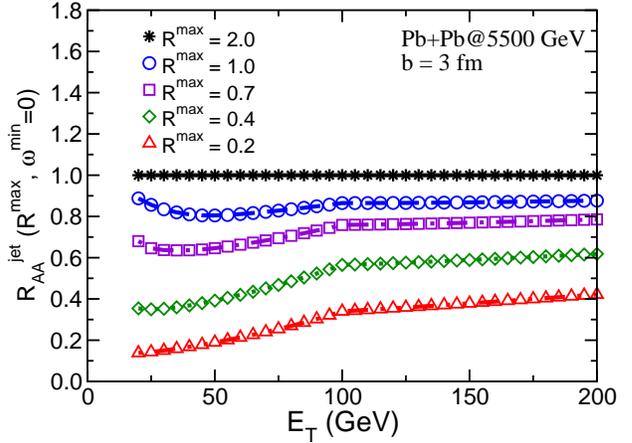}
\end{center}
\caption{ $E_T$-dependent nuclear modification factor  
$R_{AA}^{\texttt{jet}}(R^{\max},\omega^{\min})$ for different jet cone radii 
$R^{\max}$  at $b=3$~fm in central $Pb+Pb$ collisions with $\sqrt{s}=5500$~GeV.
}
\label{fig:Raa-2}
\end{figure}

Clearly, the centrality dependence of $R_{AA}^{\texttt{jet}}(R^{\max},
\omega^{\min})$ at different impact parameters will remain an 
important handle in understanding the QCD medium. But the
major development, that we report on, is the emerging understanding on how 
to get directly at the mechanisms of jet interaction in matter.
In Fig.~\ref{fig:Raa-2} we show  the evolution of  the nuclear modification 
factor for the jet cross section by varying the size of the cone radius 
$R^{\max}$ ($\omega^{\min=0}$). We see  that with increasing  cone 
radius $R^{\max}$ the quenching of the jet cross section disappears
at all $E_T$ and finally reaches unity when $R^{\max} =2.0$ and
all of the lost energy is re-captured. It is important to note the
factor of 5 to 10 variation in the quenching of the jet cross 
section. Such variation will pinpoint the angular distribution of
the bremsstrahlung gluons. We have also studied in detail the dependence
of  $R_{AA}^{\texttt{jet}}(R^{\max},\omega^{\min})$ on the acceptance 
cut $\omega^{\min}$~\cite{VWZ} and found similar sensitivity to
the gluon energy dependence of $dI^g/d\omega$. The continuous 
variation of quenching values may help differentiate between 
competing models of parton energy loss, thereby eliminating the
order of magnitude uncertainty in the extraction of the QGP density.

\begin{figure}[!t]
\vspace*{.5cm}
\begin{center}
\epsfig{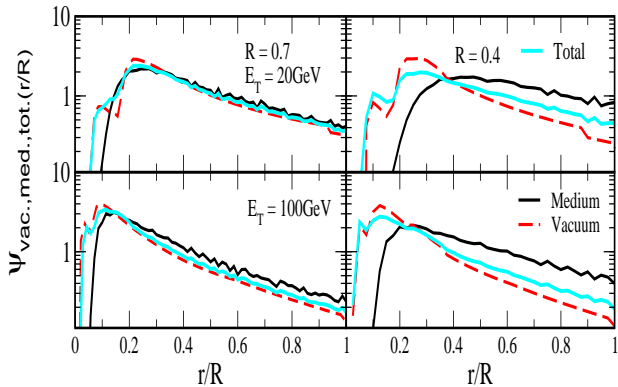} 
\end{center}
\vspace*{.2cm}
\caption{Comparisons of the jet shape in vacuum, 
medium-induced jet shape, and the total jet shape for cone radii
$R=0.7,\; 0.4$  and for different jet energies  $E_T=20,50,100,200$~GeV, respectively, 
in $Pb+Pb$ collisions at LHC.}
\label{fig:Rpsi-jet}
\end{figure}

Next, we turn to the pattern of energy flow for in-medium jets.
In Fig.~\ref{fig:Rpsi-jet} the jet shape in vacuum, the medium-induced 
jet shape and the total jet shape in the QGP  with different selection
of cone radii  and different transverse energies $E_T$ in central 
$Pb+Pb$ collisions at $\sqrt{s}=5.5$~TeV are shown together for 
comparison. An interesting conclusion is that there is no significant 
distinction  between the jet shape in vacuum and the total in-medium $\psi(r/R)$.
The underlining reason for this surprising result is that although 
medium-induced gluon radiation produces a broader $\psi_{\rm med.}(r/R)$, 
this effect is offset by the fact that the jets lose a finite
amount of their energy. Furthermore, when part of the lost energy is 
missed due to finite acceptance, the required higher initial virtuality 
jets are inherently narrower.

\begin{table}
\caption{ Summary of the mean relative jet radii  in vacuum and in the QGP medium. 
Jet cone radii $R=0.4$ and $R=0.7$ 
and energies $E_T = 20$ and $100$~GeV at $ \sqrt{s}=5.5$~TeV  central $Pb+Pb$ 
collisions at the LHC are shown. }
\label{table:mean-radii}
\begin{tabular}{ccc}
\hline\noalign{\smallskip}
   $ R=0.4 $   &  Vacuum    & Realistic case   \\
\noalign{\smallskip}\hline \noalign{\smallskip}
   $\langle r/R \rangle$, $E_T=20$GeV   &  0.41     &  0.45  \\
   $\langle r/R \rangle$, $E_T=100$GeV  &  0.28     &  0.32  \\
\noalign{\smallskip}\hline \hline\noalign{\smallskip}
   $ R=0.7 $   &  Vacuum   & Realistic case   \\ 
\noalign{\smallskip}\hline \noalign{\smallskip} 
  $\langle r/R \rangle$, $E_T=20$GeV   &  0.41      &  0.42 \\
  $\langle r/R \rangle$, $E_T=100$GeV  &  0.27     &  0.29  \\
\noalign{\smallskip}\hline
\end{tabular}
\end{table}
In Table~\ref{table:mean-radii} we present the mean relative jet radii 
 $\langle r/R \rangle$ in the vacuum and in the QGP medium created at the LHC 
for  two different cone selections $R=0.4$ and $R=0.7$  and two jet energies 
$E_T = 20, 100$~GeV. We see that in the realistic numerical simulation of quark 
and gluon propagation in the QGP  there is very little $<+10\%$ variation 
in  this observable. This  characteristic difference is slightly larger
for a smaller cone, since it emphasizes the large-angle character
of the medium-induced radiation~\cite{GLV,Vitev:2005yg}. 
It is important to stress that the QGP is rather  ``gray'' than 
``black'' and only a fraction of the energy of the leading jet is 
lost in the medium. The effect of even a moderate 
$\frac{\Delta E^{in}}{E}(R^{\max},\omega^{\min})$ can be amplified by the steeply 
falling cross sections for the $R_{AA}^{\texttt{jet}}(E_T; R^{\max},\omega^{\min})$ 
observable but this is not the case for $\langle r/R \rangle$.

\begin{figure}[!t]
\begin{center}
\epsfig{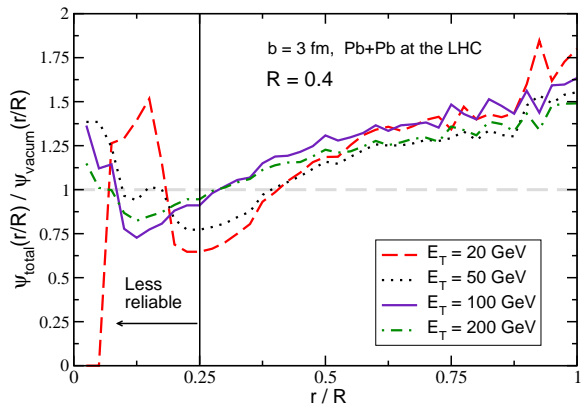} 
\end{center}
\caption{ The ratios of the total jet shape in heavy-ion collisions to the jet 
shape in vacuum for jet energies $E_T=20,50,100,200$~GeV and cone radius 
$R=0.4$ in $b=3$~fm  $Pb+Pb$ collision at $\sqrt{s}=5.5$~TeV. }
\label{fig:Rpsi-ratio}
\end{figure}
Finally, we point out where the effect of the QGP medium  can be more readily  
observed in the differential jet shapes. These are the ``tails'' of the 
energy flow distribution in $\psi_{\rm tot.}(r/R)$, near the periphery of the 
jet cone. We can see that, with cone radius $R=0.4$, the ratio  
$\psi_{\rm tot.}(r/R) / \psi_{\rm vac.}(r/R) $  could reach 
about $1.75$ when $r/R \rightarrow 1$. However, for experiments to  
reliably detect and quantify this enhancement of the ratio of  the total jet 
shape in the  QCD medium to the jet shape in the vacuum tn the $ r/R > 0.5$ region,  
high  statistics  measurements will be necessary. The required precision can 
be achieved with the large acceptance experiments at the LHC beyond the 
proof-of-principle jet identification  at RHIC~\cite{Salur:2008hs}.

\section{Conclusions}
\label{sec:summary}

The unprecedentedly high center of mass energy at the LHC will open a new
frontier for jet physics in dense QCD matter: tomography {\em of} jets that can 
help pinpoint energy flow in strongly-interacting systems  and significantly reduce  
uncertainties in the QGP properties extracted from different phenomenological 
models. In this summary we highlighted several important results from 
a recent comprehensive study of jet shapes and jet cross sections 
in $\sqrt{s}=5.5$~TeV  $Pb+Pb$ collisions at the LHC~\cite{VWZ}. 
A theory of jet shapes in the vacuum, generalized to deal with experimental 
acceptance cuts, was discussed and then applied to study jet shapes in heavy-ion 
collisions by including the contribution from the medium-induced gluon bremsstrahlung. 
Realistic numerical simulations of jet cross sections and jet shapes were used 
to illustrate new signatures of quark and gluon propagation in the QGP.  
It was shown that a suitable generalization of the nuclear modification factor 
for jets, $R_{AA}^{\texttt{jet}}(R^{\max},\omega^{\min})$, provides a sensitive 
tool to probe the energy and angular dependence of the stimulated gluon
emission pattern. The largest difference between the jet shape in 
vacuum and the total in-medium 
jet shape was manifest in the periphery of the cone, $r/R \rightarrow 1$,  and for 
smaller radii, e.g.  $R^{\max}=0.4$.  The enhancement of the mean relative 
jet radius $\langle r/R \rangle $  due to QGP effects was rather modest.

\vspace*{.3cm}

{ \bf Acknowledgments:} We thank M. H. Seymour and H. Caines for many helpful
discussions. This research is supported by the US Department
of Energy, Office of Science, under Contract No. DE-AC52-06NA25396
and in part by the LDRD program at LANL, the MOE of China 
under Project No. IRT0624 and the NNSF of China.

\end{document}